\documentclass[twocolumn,aps,prc,amsmath,showpacs,amssymb,floatfix]{revtex4-1}
\usepackage{latexsym}
\usepackage{graphicx,epsfig}
\usepackage{bm}
\usepackage{longtable}
\usepackage{mathptmx}                
\usepackage{dcolumn}                 

\def\rra{\right\rangle}
\def\lla{\left\langle}

\def\g{$g$-mode~}
\def\sg{$g$-modes~}
\def\f{$f$-mode~}
\def\msun{M_\odot}
\def\ii{{\rm i}}

\def\eps{\epsilon}
\def\om{\omega}
\def\lm{{\ell m}}
\def\lgw{L_\text{GW}}

\begin{document}

\title{Oscillations of hot, young neutron stars:\\
Gravitational wave frequencies and damping times}

\author{
G.~F.~Burgio,$^1$ V.~Ferrari,$^2$ L.~Gualtieri,$^2$ and H.-J.~Schulze$^1$}

\affiliation{
$^1$INFN, Sezione di Catania, Dipartimento di Fisica,
Via Santa Sofia 64, 95123 Catania, Italy\\
$^2$Dipartimento di Fisica, ``Sapienza'' Universit\`a di Roma
\& Sezione INFN Roma1, P.A.~Moro 5, 00185 Roma, Italy}

\begin{abstract} 
We study  how the
frequencies and damping times of oscillations  of a newly born,
hot proto-neutron star depend on the physical quantities which characterize the
star quasi-stationary evolution which follows the bounce.
Stellar configurations are modeled using a microscopic equation of state
obtained within the Brueckner-Hartree-Fock,  nuclear many-body
approach, extended
to the finite-temperature regime. We discuss the mode frequency
behaviour as function of the lepton composition, and
of the entropy gradients which prevail in the interior of the star.
We find that, in the very early stages, gravitational wave emission 
efficiently competes with neutrino processes in dissipating the star
mechanical energy residual of the gravitational collapse.

\end{abstract}

\pacs{
04.30.Db, %
97.10.Sj, %
97.60.Jd, %
26.60.Kp  %
}

\maketitle

\section{Introduction}
\label{s:intro}
The birth of  a proto-neutron star (PNS) in a core collapse supernova
is a very difficult phenomenon to model, since it requires not only
accurate descriptions of the micro-physics of the collapsing matter,
in particular of neutrino transport and related processes, but also
of the violent dynamical processes occurring in the
contracting-exploding star, which need to be treated in the framework of
general relativity (see \cite{ott2009} for a recent review).

The description of the subsequent PNS evolution is also
challenging, because a PNS is a hot and rapidly evolving object.  The
physical processes which contribute to the star cooling and contraction,
such as nuclear and weak interactions and energy and lepton
number transport by neutrino diffusion, have to be included in 
dynamical simulations.  Thus, most simulations of
gravitational core collapse to a PNS end shortly after the core bounce
and the launch of the supernova explosion -- typically after
a few hundreds of milliseconds --
and only a few dynamical simulations extend to the first
minute of the PNS life \cite{pons1,pons2,pons3,Mezzacappaetal_2010}.

In this paper we are interested in this latest phase of the PNS life,
when shock waves, neutrino winds, convection instabilities, and
accretion flows are no longer dominant and the star cooling and
contraction proceed on timescales of seconds, so that the evolution is
quasi-stationary.  In particular, we want to study how 
the frequencies and damping times of the PNS quasi-normal modes of
oscillation depend on the star internal structure.  The main motivation 
is that  the oscillations  of a newly born PNS may be associated to
gravitational wave signals  with sizeable 
amplitudes, and with frequencies lower than those typical of mature neutron
stars. This would favour  their detection by the next generation
of ground-based interferometers LIGO/Virgo and their future version (ET)
\cite{ferrari_miniutti_pons_2003,reviewmia,ET}.

The available dynamical simulations of the post-bounce evolution of a
PNS indicate that, typically, the star goes through the following
main steps.  After the core bounce, a shock wave propagates through
the outer PNS mantle, leaving behind a low-entropy core in which
neutrinos are trapped, surrounded by a low-density, high-entropy
envelope.  The mantle accretes matter from the outer layers and
rapidly contracts, losing energy due to electron captures and thermal
neutrino emission.  The supernova explosion lifts off the stellar
envelope and, in a few tenths of seconds, due to extensive neutrino
losses, the lepton pressure decreases and the envelope contracts.  At
this stage the PNS radius is about 20--30~km;  the subsequent
evolution can be described as a sequence of equilibrium
configurations; this quasi-stationary evolution is the phase of interest for us.

Simulations show that the diffusion of high-energy neutrinos
(of the order of a few hundred MeV) from the low-entropy core 
to the high-entropy envelope, from which they finally escape
with energies of the order of a few tens of MeV,  generates
a large amount of heat within the star, producing temperatures up
to several tens of MeV; as a result, the core entropy approximately
doubles, whereas the entropy of the  envelope decreases.
In a few tens of seconds the PNS becomes lepton poor, but 
it is still hot.  The net number of neutrinos in the interior is low,
but thermally produced neutrino pairs of all flavors are abundant, and
dominate the emission; the star cools down and
entropy gradients are gradually smoothed out,
while the average neutrino energy decreases, and neutrinos
mean free path increases;  after approximately one minute  it
becomes comparable to the stellar radius, and the star
becomes neutrino transparent.  By this time, the temperature has
dropped to 1--5~MeV ($\approx$ 1--5$\times 10^{10}$~K) and the star
has radiated off almost all of its binding energy, becoming what we
call a neutron star (NS).

This brief summary of the first minute of the PNS life is deliberately
imprecise, because the details of the evolution depend on the
assumptions on which the simulation is based.  For instance, in
\cite{pons1,pons2,pons3}, where the first minute after the core bounce
is considered, the evolution is treated as a sequence of
quasi-stationary states: the thermodynamical variables and the lepton
fractions are determined by solving evolution equations (for instance
Boltzmann's equation to model neutrino transport), whereas at each
time-step the stellar structure is found by solving the
Tolman-Oppenheimer-Volkov equations.  In \cite{Mezzacappaetal_2010},
instead, all quantities are determined through a time-evolution
core-collapse code, which has been extended in order to describe the
first $\sim 20$ s of the post-bounce processes. 
However, there are features
which are common to different studies; indeed,
starting from an initial configuration characterized by a low-entropy core
and a high-entropy envelope, due to neutrino processes the star goes
smoothly through the following phases:
\begin{itemize}
\item
the entropy increases in the core  while decreasing in  the envelope;
\item
entropy gradients gradually smooth out while the star is still hot;
\item
the star progressively cools down, and the overall entropy decreases;
\item
the evolution ends in the ``cold'', zero entropy,  neutron star configuration. 
\end{itemize}
We remark that both
in \cite{pons1,pons2,pons3} and in \cite{Mezzacappaetal_2010}
the equation of state (EOS) of
baryonic matter is a finite-temperature, field-theoretical model
solved at the mean field level.

In this paper, instead, we employ a microscopic EOS obtained within
the Brueckner-Hartree-Fock (BHF) nuclear many-body approach extended
to the finite temperature regime, as we shall discuss in
Sec.~\ref{ss:eos}. We stress  that our aim is not to model
the PNS quasi-stationary evolution;  we rather want to explore how the
frequencies and damping times of the PNS quasi-normal modes 
depend on the physical quantities which characterize the
quasi-stationary configurations, which are essentially the entropy
profile (which will appear to be the most important in this respect)
and the lepton composition. 

PNS are expected to be rapidly rotating; however in our study we neglect 
rotation, since we are primarily interested in 
the effects of the thermal and chemical evolution on the star oscillation
frequencies, and in comparing the results with those of previous works
which use different EOSs to model the PNS. 

The article is organized as follows.  In Sec.~\ref{s:osc} we briefly
explain how to compute  the complex values of the quasi-normal mode frequencies
using the relativistic theory of stellar perturbations.  In
Sec.~\ref{ss:eos} the derivation of  the equation of state of 
hot nuclear matter used to model the PNS evolution
is shortly illustrated. In Sec.~\ref{prof} we discuss how the different
stages of a PNS quasi-stationary evolution are simulated by constructing
stellar configurations with appropriate entropy and lepton fraction
profiles.
In Sec.~\ref{s:res} we compute and discuss the stellar parameters and 
the  quasi-normal mode  frequencies for the various
configurations. Conclusions are drawn in Sec.~\ref{s:end}.

\section{The quasi-normal modes of neutron stars}
\label{s:osc}

\subsection{Stellar perturbations}

In order to find frequencies and damping times of the quasi-normal
modes (QNMs) of a star, we need to solve the equations describing non-radial
perturbations of a (spherically symmetric) star in general relativity,
which we briefly recall.

The perturbed spacetime metric is expanded in tensor spherical
harmonics, as (we use geometrized units, assuming $c=G=1$)
\begin{eqnarray}
  ds^2 &=& 
  -e^\psi \left( 1 + r^\ell H_0^\lm Y_\lm e^{\ii\om t} \right) dt^2
\nonumber\\&&
  +e^\lambda \left( 1 - r^\ell H_2^\lm Y_\lm e^{\ii\om t} \right) dr^2
\nonumber\\&& 
  - 2\ii\om r^{\ell+1} H_1^\lm Y_\lm e^{\ii\om t} dtdr
\nonumber\\&& 
  + r^2 \left( 1-r^\ell K^\lm Y_\lm e^{\ii\om t} \right)
  (d\vartheta^2 + \sin^2\theta d\varphi^2) \:, \quad
\end{eqnarray}
where $\om$ is the frequency, $Y_\lm(\vartheta,\!\varphi)$ are the
scalar spherical harmonics, and $H_i^\lm(r)$, $K^\lm(r)$ describe the
metric perturbations with polar parity, i.e., those transforming as
$(-1)^\ell$ under a parity transformation.  In this paper we do not
consider perturbations with axial parity, which transform as
$(-1)^{\ell+1}$.  The functions $\psi(r),\lambda(r)$ describe the
unperturbed metric, and are found by solving the
Tolman-Oppenheimer-Volkov equations.  The four-velocity of the generic
fluid element  is
\begin{equation}
 u^\mu = u_0^\mu + \delta u^\mu 
 = (e^{-\psi/2},0,0,0) + \ii\om e^{-\psi/2} (0,\xi_r,\xi_\theta,\xi_\phi) \:,
\end{equation}
where $\xi_\mu$ is the fluid element Lagrangian displacement,
expanded in vector spherical harmonics 
\begin{eqnarray}
  \xi_r(t,r,\vartheta,\varphi) &=& 
  e^{\lambda/2} r^{\ell-1} W^\lm(r) 
  Y_\lm(\vartheta,\varphi) e^{\ii\om t}
\:,\nonumber\\
  \xi_\vartheta(t,r,\vartheta,\varphi) &=& 
  -r^{\ell} V^\lm(r) \partial_\vartheta 
  Y_\lm(\vartheta,\varphi) e^{\ii\om t} 
\:,\nonumber\\
  \xi_\varphi(t,r,\vartheta,\varphi) &=& 
  -r^{\ell} V^\lm(r) \partial_\varphi 
  Y_\lm(\vartheta,\varphi)e^{\ii\om t} \:.
\end{eqnarray}

The fluid is also characterized by its pressure and energy density
\begin{eqnarray}
  p(r) + \delta p(t,r,\vartheta,\varphi) &=& 
  p(r) + r^\ell\delta p^\lm(r)Y_\lm(\vartheta,\varphi)e^{\ii\om t} \:,
\nonumber\\
  \eps(r) + \delta\eps(t,r,\vartheta,\varphi) &=& 
  \eps(r) + r^\ell\delta\eps^\lm(r)Y_\lm
  (\vartheta,\varphi)e^{\ii\om t} \:. \quad 
\end{eqnarray}
We denote with $\delta$ the Eulerian perturbations, and with $\Delta$
the Lagrangian perturbations, so that for instance the Lagrangian
perturbation of the pressure is 
\begin{equation}
  \Delta p = \delta p + \xi^r {\partial p \over \partial r} \,,
\end{equation}
i.e.,
\begin{equation}
  \Delta p^\lm = \delta p^\lm 
  + \frac{e^{-\lambda/2}}{r} W^\lm {\partial p \over \partial r} \,.
\end{equation}
Einstein's equations, linearized in the perturbations, yield a system
of ordinary differential equations for the perturbed functions.
Different equivalent sets of equations have been derived in the
literature, using different gauge choices or different combinations of
the relevant equations \cite{T67,LD83,DL85,CF}.  In this paper we use the
formulation of Lindblom and Detweiler \cite{LD83,DL85}, consisting of
a system of four first-order differential equations (hereafter, the LD
equations) for the functions
$\{H_1^\lm,K^\lm,W^\lm,X^\lm\}$, where
\begin{equation}
  X^\lm = -e^{\psi/2}\Delta p^\lm \:,
\label{defX}
\end{equation} 
and algebraic relations which allow to compute the remaining functions
$\{H_0^\lm,H_2^\lm,V^\lm\}$ in terms of the others (see Appendix
\ref{LD}).  To close the system, an EOS,
relating the energy density $\eps$ and the pressure $p$, has to be
assigned.

\subsection{The quasi-normal mode frequencies}

A QNM is a solution of the perturbation equations, which is regular at the
center, continuous on the surface, and which behaves as a pure
outgoing wave at infinity.  Since in general relativity a non-radial
oscillation is associated to gravitational wave emission,
such solutions belong to complex frequencies:
\begin{equation}
  \om = \sigma + \frac{\ii}{\tau_\text{GW}} \:,
\end{equation}
where $\sigma=2\pi\nu$  and $\nu$ is the pulsation frequency;  $\tau_\text{GW}$ is the
damping time of the mode due to gravitational wave emission.  If the
mode is unstable, its imaginary part is negative and $-\tau_\text{GW}$
is the growth time of the instability.

The procedure to find the QNM frequencies is the following: 
(i) We choose a value of $l$ and a complex value of $\om$ (since the
background is spherical, the equations do not depend on the index
$m$).
(ii) We integrate  Eqs.~(\ref{LLDD}), by imposing that the solution is regular
at the center and that $\Delta p=0$ at the stellar surface [Eqs.~(\ref{bc1}),(\ref{bc2})].
(iii) We impose that the solution and its first derivative are continuous 
on the stellar surface, and find the metric perturbations outside the star.
(iv) In vacuum, the perturbed equations reduce to a simple,
second-order differential equation (the Zerilli equation (\ref{zereq})), which we
integrate up to radial infinity.
(v) We check whether the solution satisfies the outgoing wave boundary
condition at infinity (\ref{bc3}) which identifies a quasi-normal
mode; we then repeat the procedure for different values of $\om$.  The
values of $\om$ which satisfy the outgoing wave condition can be found
using a Newton-Raphson method.

The polar QNMs are classified, following a scheme introduced by
T.G. Cowling in Newtonian theory \cite{Cox}, on the basis of the
restoring force which prevails when the generic fluid element is
displaced from the equilibrium position.  Thus, we have a $g$-mode if
the restoring force is mainly provided by buoyancy or a $p$-mode if it
is due to a pressure gradient.  The frequencies of the $g$-modes are
lower than those of the $p$-modes, and the two sets are separated by
the frequency of the fundamental ($f$-) mode, which is related to a
global oscillation of the fluid.  In general relativity there exist
further modes, named $w$-modes \cite{wmodes}, that are purely
gravitational, since they barely excite the fluid motion.  Other
classes of modes are associated to NS features which are not included
in the present model, like rotation, magnetic fields, the crust rigidity.

\subsection{Sound speed}

A neutron star at the end of its evolution is cold and isentropic,
matter is in beta-equilibrium and can be described by a barotropic EOS
$p=p(\eps)$.  Conversely, when the star is young and hot the EOS
cannot be expressed in a barotropic form, since the pressure depends
non-trivially on the entropy and on the composition, i.e.,
\begin{equation}
  p = p(\eps,s,x_i) \:.
\label{eosgen}
\end{equation}
Therefore, to solve the perturbed equations the profiles of entropy
and particle fractions, respectively $s(n)$ and $x_i(n)$, are also
needed. In Eq.~(\ref{eosgen}) $n$ is the baryon number density, $s=S/A$
is the entropy per baryon, and $x_i=n_i/n$ is the fraction of the $i$-th
particle.  Usually matter is locally in beta equilibrium and neutrinos
are trapped, therefore the dependence on the composition $\{x_i\}$
reduces to a dependence on the lepton fraction $Y_e=x_e+x_{\nu_e}$
only.

The perturbed equations (\ref{LLDD})  depend explicitly on the sound speed $c_s^2$,
which relates the Lagrangian perturbations of pressure and energy
density,
\begin{equation}
  \Delta p = c_s^2 \Delta\eps \:.
\label{dpcs2de}
\end{equation}
$c_s^2$ is defined as the following thermodynamical derivative
\begin{equation}
  c_s^2 = \left(\frac{\partial p}{\partial\eps}\right)_\text{adiabatic} \:,
\label{dpcs2dea}
\end{equation}
where ``adiabatic'' means that the derivative is performed keeping
fixed the entropy and the fractions of those particle species {\em
  which do not change during the pulsation} \cite{MT66}.

To clarify this statement, let us consider a fluid element oscillating
with period $t_\text{osc}$ about the equilibrium position.  The
following equation holds:
\begin{equation}
  \Delta p = \left(\frac{\partial p}{\partial\eps}\right)_{s,x_i}
  \!\!\!\!\Delta\eps
  + \left(\frac{\partial p}{\partial s}\right)_{\eps,x_i}
  \!\!\!\!\Delta s
  + \sum_i\left(\frac{\partial p}{\partial x_i}\right)_{\eps,s}
  \!\!\!\!\Delta x_i \:.
\label{dpall}
\end{equation}
Since we are considering adiabatic perturbations, the fluid element
does not exchange heat with its surroundings and $\Delta s=0$.
Furthermore, the displaced fluid element has a composition different
from the surrounding fluid even though nuclear reactions, acting on a
timescale $t_\text{react}$, tend to eliminate this difference.  The
two limiting cases are:

i) $t_\text{react} \gg t_\text{osc}$; in this case the 
fluid element composition does not change during the oscillation,
i.e., $\Delta x_i=0$, and by combining Eqs.~(\ref{dpcs2de}) and
(\ref{dpall}) we find
\begin{equation}
  c_s^2 = \left(\frac{\partial p}{\partial\eps}\right)_{s,x_i}\:.
\end{equation}

ii) $t_\text{react} \ll t_\text{osc}$;  the fluid element
composition changes, becoming that of the surrounding fluid.  By
replacing the composition profile $x_i=x_i(\eps,s,Y_e)$ in
(\ref{eosgen}), it is possible to express the EOS as
$p=p(\eps,s,Y_e)$.  Eqs.~(\ref{dpcs2de}) and (\ref{dpall}) then give
\begin{equation}
  c_s^2 = \left(\frac{\partial p}{\partial\eps}\right)_{s,Y_e} \:.
\end{equation}
for beta-stable,  neutrino-trapped matter.

For the PNSs in quasi-stationary evolution we consider in this paper,
typical oscillation  periods are of the order $t_\text{osc} \approx 10^{-3}$ s,
while weak interactions timescales are \cite{HLY02}
\begin{equation}
  t_\text{react}^{(1)} \approx \frac{5\times 10^6\;\text{s}}{(T/10^9K)^6} \:,
~~~~
  t_\text{react}^{(2)} \approx \frac{20\;\text{s}}{(T/10^9K)^4}
\end{equation}
for modified and direct Urca processes, respectively.  In the first
seconds of a PNS life $T \approx (1-4) \times 10^{11}$ K, thus
$t_\text{react}^{(1,2)} \ll t_\text{osc}$.  Therefore the stellar
pulsations always occur in local beta equilibrium.  Timescales of
strong nuclear reactions are even smaller.

\subsection{Some considerations on the $f$- and $g$-modes}
\label{s:gmodes}

For old, cold neutron stars, the frequency of the $f$-mode, $\nu_f$,
is in the range 1-3 kHz, and the damping time, $\tau_f$, is of the
order of a few tenths of seconds.  According to the Newtonian theory
of stellar pulsations, $\nu_f$ scales as the square root of the
star average density, and this behaviour is maintained in the
relativistic theory, according to which the damping time scales as
$\tau_f \sim R^4/M^3$ \cite{asterosism0,asterosism,chandra}.  

The \sg are directly related to the thermodynamical properties of the
star.  Indeed, their presence can be traced back to the Schwarzschild
discriminant \cite{T66},
\begin{equation}
  S(r) = \frac{dp}{dr}-c_s^2\frac{d\eps}{dr} =
  \frac{dp}{dr}\left(1-\frac{c_s^2}{c_{s0}^2}\right) \,,
\end{equation}
where $c_{s0}^2 = \frac{dp/dr}{d\eps/dr}$. 
The radial acceleration of a fluid element
displaced from equilibrium by $\Delta r$ is
\begin{equation}
a=-\frac{e^{-\lambda/2}}{(\epsilon+p)^2c_s^2}\left|\frac{dp}{dr}\right|S(r)\Delta r\,.
\end{equation}
Therefore, if $S(r)>0$ the fluid element oscillates about the
equilibrium position, whereas if $S(r)<0$ it is accelerated away from
equilibrium.
It follows that, if in some region of the star $S(r)<0$, 
this region is convectively unstable and the star admits a set of
unstable $g$-modes, otherwise the $g$-modes are stable.
If $S(r)$ vanishes identically, the star does not have \sg (all \sg
degenerate to zero frequency).  This is the case if the neutron star
is cold and old, since the EOS is barotropic and
$c_s^2 = \frac{p_{,r}}{\eps_{,r}}$.
Similar information is contained in the Brunt-V\"ais\"al\"a frequency
which, in a relativistic framework, is defined as \cite{MVS83}
\begin{equation}
  N^2(r) = \frac{e^{\psi-\lambda}}{(\eps+p)c_s^2}
  \frac{\psi_{,r}}{2} S(r) \:.
\end{equation}
It has been shown that, although the
Brunt-V\"ais\"al\"a frequency changes by many
orders of magnitude throughout the star, it allows to estimate some \g
frequencies of Newtonian stars.  For instance, in \cite{T80} the
frequency of higher-order \sg of main sequence stars is computed using
the following formula
\begin{equation}
  \sigma_{g} \approx \frac{\sqrt{\ell(\ell+1)}}
  {\left(2\kappa+\ell+n_e+\frac{1}{2}\right)\frac{\pi}{2}}
  \int_0^R dr \frac{|N(r)|}{r} \:,
\label{tassg}
\end{equation}
where $R$ is the stellar radius, $\kappa$ is the order of the
$g$-mode, and $n_e$ is the effective polytropic index of the outer
layers of the star.  However, the Brunt-V\"ais\"al\"a frequency cannot
be used to estimate neutron star $g$-mode frequencies \cite{MVS83};
these frequencies can only be found by solving the perturbation
equations, as we do in the present paper.  Nevertheless the following
considerations will be useful to interpret the results we will show in
the following.  Eq.~(\ref{tassg}) indicates that higher frequency
values correspond to larger values of $S(r)$ (i.e., of $|N(r)|$)
inside the star.  Since we shall assume $dY_e/dr=0$ (see
Sec. \ref{prof}), we have
\begin{equation}
  S(r) = \frac{dp}{dr} - 
  \left(\frac{\partial p}{\partial\eps}\right)_{s,Y_e}\frac{d\eps}{dr} =
  \left(\frac{\partial p}{\partial s}\right)_{\eps,Y_e}\frac{ds}{dr} \:,
\end{equation}
we may expect that higher $g$-mode frequencies correspond to
larger entropy gradients.

\subsection{The damping time of quasi-normal modes}
\label{s:tau}

A QNM of a PNS is characterized by the pulsation frequency and
by the damping time $\tau_\text{GW}$.  Its value is important
because it shows how fast the pulsation energy can be dissipated
through gravitational wave emission, and it must be compared to the
timescale $\tau_\text{diss}$ associated with other dissipative
processes which may compete with GW emission.  These include
viscosity, heat transport, neutrino diffusion, etc.  (phenomena which
we are neglecting in our model).  In the first minute of life of a
PNS, $\tau_\text{diss}\sim10-20$ s \cite{GP82,HW84} (see also the
discussion in \cite{ferrari_miniutti_pons_2003}). Thus, if $\tau_\text{GW}
\ll\tau_\text{diss}$, the oscillations are mainly damped by
gravitational wave emission, and viceversa.  We also remark that, if a
QNM is unstable, the instability can grow only if $\tau_\text{GW} \ll
\tau_\text{diss}$.

As long as $\tau_\text{GW} \ll \tau_\text{diss}$, when a star
oscillates in a QNM, the pulsation energy changes in time as \cite{T67}
\begin{equation}
  E_\text{puls}(t) \approx E_\text{puls}(0) e^{-2t/\tau_\text{GW}},
\end{equation}
and the power radiated in gravitational waves is
\begin{equation}
  \lgw = -\dot E_\text{puls} \approx 2E_\text{puls}/\tau_\text{GW} \:.
 \label{lepuls}
\end{equation}
Thus, smaller QNM damping times are associated with a more efficient
gravitational wave emission.  In the case of cold NSs, the \f has
always the smallest damping time, but this is not always the case for
PNSs,  as we shall discuss later.

Although we shall compute the damping times of all modes by direct
integration of the perturbed equations, it is useful to give some
approximate formula which will allow us to explain some results of
the next sections.  From Eq.~(\ref{lepuls}) we find
\begin{equation}
  \tau_\text{GW} \approx 2E_\text{puls}/\lgw \;.
\label{e:estimtau}
\end{equation}
The (approximate) expressions of $E_\text{puls}$ and $\lgw$ (the
latter is obtained using the quadrupole formalism) can be found in
\cite{T69,MVS83} and are, in terms of the perturbation functions
defined in this paper,
\begin{eqnarray}
  E_\text{puls} &\approx& \frac{1}{2}\sigma^2
  \int_0^R dr~r^{2\ell} (\eps+p) e^{(\lambda-\psi)/2}
\nonumber\\ && \times
  \left[ |W^\lm|^2 + \ell(\ell+1) |V^\lm|^2 \right]
\end{eqnarray}
and
\begin{equation}
  \lgw \approx \frac{4\pi}{75}\sigma^6
  \left| \int_0^R dr~r^4 \delta\eps^\lm \right|^2 \:,
\label{LGW}
\end{equation}
where 
\begin{equation}
  \delta\eps^\lm = -r^\ell \left[
  \frac{e^{-\psi/2}}{c_s^2}X^\lm +
  \eps_{,r}\frac{e^{-\lambda/2}}{r}W^\lm \right] \:.
\label{e:deltaeps} 
\end{equation}

\section{The equation of state of hot nuclear matter}
\label{ss:eos}
\subsection{BHF many-body approach}
\label{BHFap}
One of the most advanced microscopic approaches to the EOS of nuclear
matter is the Brueckner theory \cite{book}, recently extended to the
finite-temperature regime within the Bloch-De Dominicis formalism
\cite{bloch}.  In this approach, the essential ingredient is the
two-body scattering matrix $K$, which, along with the single-particle
potential $U$, satisfies the self-consistent equations
\begin{eqnarray}
 && \lla 1 2 | K(W) | 3 4 \rra
 = \lla 1 2 | V | 3 4 \rra  
\nonumber\\ 
 && +\; \mathrm{Re} \sum_{3',4'} 
 \langle 1 2 | V | 3' 4' \rangle 
 { [1-n^F(3')] [1-n^F(4')] \over W - E_{3'} - E_{4'} + i\varepsilon }
 \langle 3' 4' | K(W) | 3 4 \rangle \quad
\nonumber\\&&
\label{eq:kkk}
\end{eqnarray}
and
\begin{equation}
 U(1) = \sum_{2} n^F(2) \lla 1 2 | K(W) | 1 2 \rra_A \:,
\label{eq:ueq}
\end{equation}
where $1,2,...$ generally denote momentum, spin, and isospin.  Here
$V$ is the two-body interaction, $W = E_{1} + E_{2}$ represents the
starting energy, and $E_i = k_i^2/2m_i + U(k_i)$ the single-particle
energy; $n^F(k)$ is the Fermi distribution at finite temperature.  For
assigned partial densities and temperature, Eqs.~(\ref{eq:kkk}) and
(\ref{eq:ueq}) have to be solved self-consistently along with the
following equations for the auxiliary chemical potentials
${\tilde{\mu}}_i$,
\begin{equation}
 n_i = \sum_k n^F_i(k) = 
 \sum_k {1\over e^{\beta (E_i(k) - {\tilde{\mu}}_i)} + 1 } \:,
\label{e:ro}
\end{equation}
and the baryon number density is $n=\sum_in_i$.

At finite temperature the EOS, and all thermodynamical
quantities, can be obtained from the grand-canonical potential density
$\om$.  In the Bloch-De Dominicis approach, $\om$ can be written as
the sum of a mean-field term and a correlation contribution
\cite{baldo,book},
\begin{eqnarray}
 \om &=& -\sum_k \left[ {1\over\beta} 
 \ln\left( 1 +  e^{-\beta (E_k - \tilde{\mu})}\right) 
 + n^F(k) U(k) \right]
\nonumber\\ && +  
 {\frac{1}{2}} \sum_{k} n^F(k) U(k) \:.
\label{e:om1}
\end{eqnarray}
In this framework, the free energy density is 
\begin{equation}
 f = \om + n \tilde{\mu} \:,
\label{e:f}
\end{equation}
and all remaining thermodynamical quantities of interest, namely, the
``true" chemical potential $\mu$, pressure $p$, entropy per baryon $s$,
and energy density $\epsilon$ can be computed from it as
\begin{eqnarray}
 \mu &=& {{\partial f}\over{\partial n}} \:,
\\
 p &=& n^2 {{\partial (f/n)}\over{\partial n} }  
 = \mu n - f \:,
\\
 s &=& -\frac{1}{n}{{\partial f}\over{\partial T}} \:,
\\
 \eps &=& f + Tns + m_nn \:
\end{eqnarray}  
($m_n$ neutron mass).  Since at zero temperature the non-relativistic
microscopic approaches do not correctly reproduce the nuclear matter
saturation point, $n_0 \approx 0.17~\mathrm{fm}^{-3}$, $E/A \approx
-16$ MeV, three-body forces (TBF) among nucleons are usually
introduced in order to correct this deficiency.  Given the current
lack of a complete microscopic theory of TBF, we have adopted the
phenomenological Urbana model \cite{uix}, which consists of an
attractive term due to two-pion exchange with excitation of an
intermediate $\Delta$ resonance and a repulsive phenomenological
central term.  For simplicity, we reduce the TBF to a
density-dependent two-body force by averaging over the position of the
third particle, assuming that the probability of having two particles
at a given distance is reduced according to the two-body correlation
function.  The corresponding EOS at zero temperature reproduces the
nuclear matter saturation point correctly \cite{bbb,zhou,zhli}, and
fulfills several requirements from the nuclear phenomenology.  In all
calculations presented in this paper we use the Argonne $V_{18}$
nucleon-nucleon potential \cite{v18} together with the
phenomenological Urbana TBF.

Results for symmetric nuclear matter and purely neutron matter have
been obtained for different values of the temperature, and are
discussed in \cite{nbbs,isen1,isen2,hypt}.  In particular, in
Ref.~\cite{isen2} useful numerical parametrizations of the EOS are
given that are employed in the current work.

The Brueckner approach provides a realistic modeling of nuclear matter
only at densities above about half normal nuclear matter density.
Below this threshold, clusterization sets in, and the system becomes
inhomogeneous.  Therefore, in this ``low-density'' regime
another theoretical approach  has to be used,
and we employ the EOS of Shen
\cite{shen}, which is essentially a liquid-drop-type model at finite
temperature.

Of course, since two different theoretical descriptions of the same
state of matter are involved, the joining of the two EOSs requires the
thermodynamical observables to be continuous functions of the baryon
density.  In practice we perform a Maxwell construction by equating
pressure and chemical potentials of the low- and high-density sectors,
and verify that the other thermodynamic variables do not exhibit
significant discontinuities at the transition point.  In this way, a
very wide range of baryon density is spanned, from the iron density at
the surface up to 8-10 times the nuclear saturation density in the
core.  Further details are discussed in the following subsection.

\subsection{Composition and EOS of hot stellar matter}\label{subscomp}

In neutrino-trapped $\beta$-stable nuclear matter, the chemical
potential of any particle $i=n,p,l$ is uniquely determined by the
conserved quantities, baryon number $B_i$, electric charge $Q_i$, and
weak charges (lepton numbers) $L^{(e)}_i$, $L^{(\mu)}_i$:
\begin{equation}
  \mu_i = B_i\mu_n - Q_i(\mu_n-\mu_p)
  + L^{(e)}_i\mu_{\nu_e}  + L^{(\mu)}_i\mu_{\nu_\mu} \:.
\label{e:mufre}
\end{equation}
For stellar matter containing nucleons and leptons as relevant degrees
of freedom, the chemical equilibrium conditions read explicitly as
\begin{equation}
 \mu_n - \mu_p = \mu_e - \mu_{\nu_e} = \mu_\mu + \mu_{\bar{\nu}_\mu} \:.
\label{e:beta}
\end{equation}
At given baryon number density $n$, these equations have to be solved
together with the charge neutrality condition
\begin{equation}
  \sum_i Q_i x_i = 0,
\label{e:neutral}
\end{equation}
and those expressing conservation of lepton numbers
\begin{equation}
  Y_l = x_l - x_{\bar l} + x_{\nu_l} - x_{\bar{\nu}_l}
  \:,\quad l=e,\mu \:.
\label{e:lepfrac}
\end{equation}
When neutrinos have left the system, their partial densities and
chemical potentials vanish and the above equations simplify
accordingly.  We fix the muon fractions to $Y_\mu=0$, and let $Y_e$
assume a finite value different from zero in neutrino-trapped matter.

The nucleon chemical potentials are obtained from the free energy
density $f$, Eq.~(\ref{e:f}),
\begin{eqnarray}
 \mu_i(\{n_j\}) &=&
 \left. \frac{\partial f}{\partial n_i} \right|_{n_{j\neq i}} \:,
 \ i=n,p \:,
\label{mun:eps}
\end{eqnarray}
and the chemical potentials of the non-interacting leptons are
obtained by solving numerically the free Fermi gas model at finite
temperature.  Once the hadronic and leptonic chemical potentials are
known, one can proceed to calculate the composition of the
$\beta$-stable stellar matter, and then the total pressure $p$ through
the usual thermodynamical relation
\begin{equation}
 p = n^2 {\partial{(f/n)}\over \partial{n}}
 = \sum_i \mu_i n_i - f  \:.
\end{equation}

An important feature of the low-density domain is the treatment of
neutrino trapping.  Physically, neutrinos escape rapidly from the
low-density matter during the PNS evolution, and the lepton number is
not conserved anymore.  This effect can be roughly modeled by
introducing a neutrino-sphere inside which neutrinos are trapped.
Typical model-dependent values for the location of the neutrino-sphere
found in the literature are $2\times10^{-3}\; \rm fm^{-3}$
\cite{gondek}, $6\times10^{-4}\; \rm fm^{-3}$ \cite{stro}, and
$2\times10^{-5}\; \rm fm^{-3}$ \cite{fischer}.  Given these
variations, we choose the following ``natural cutoff'' procedure: when
imposing a constant $Y_e$ at any density, at a certain threshold
number density $n_\nu \approx 10^{-5}-10^{-6}\; \rm fm^{-3}$, the electron
fraction $x_e$ becomes equal to $Y_e$, and neutrinos disappear
naturally.  For lower densities we consider the matter untrapped.
This simple procedure avoids making assumptions about the
neutrino-sphere, although a more satisfactory treatment of neutrino
trapping is required; but this is beyond the main goal of the present
paper.

\section{Proto-neutron star stellar models:
 entropy and lepton fraction profiles}
\label{prof}

We shall now construct equilibrium stellar models, all with a fixed
baryonic mass $M_B=1.5\,\msun$ (a conserved quantity during the
stellar evolution), and with different entropy and composition
profiles.  These configurations will be used to simulate the
quasi-stationary evolution of a PNS, and to compute how the stellar
parameters and the quasi-normal mode frequencies change during the
evolution.  
As discussed in the introduction, the quasi-stationary evolution
typically starts with  configurations characterized by 
a low entropy per baryon in the core (order of $s\sim 1$ at the
center, see for instance \cite{pons1}) and a large entropy per baryon
in the envelope (order of $s\sim 5$ or larger).
Thus, we shall consider as  ``initial'' the configuration
with an  entropy per baryon profile  made of
two constant pieces, $s_c =1$ in the core and $s_e=5$ in the envelope, with
a smooth junction between them.  Furthermore, as discussed in
Sec.~\ref{subscomp},  to model neutrino trapping we shall assume
that the lepton fraction $Y_e$ is constant throughout the star (up to
a threshold density, below which $Y_e=x_e$).

As long as the evolution proceeds, entropy gradients are gradually
smoothed out: the core entropy increases, the envelope entropy
decreases, neutrinos escape from the surface and the star
progressively cools down.  
To model this evolution, we construct EOSs and stellar configurations
corresponding to increasing values of $s_c$, decreasing values of
$s_e$ and decreasing lepton fraction; then, to decreasing values of
both $s_c$, $s_e$, and decreasing lepton fraction.  Each configuration
depends on the three constants $s_c,s_e,Y_e$, and it is labeled by
$P_{s_c,s_e,Y_e}$.

\begin{figure}[t]
\epsfig{file=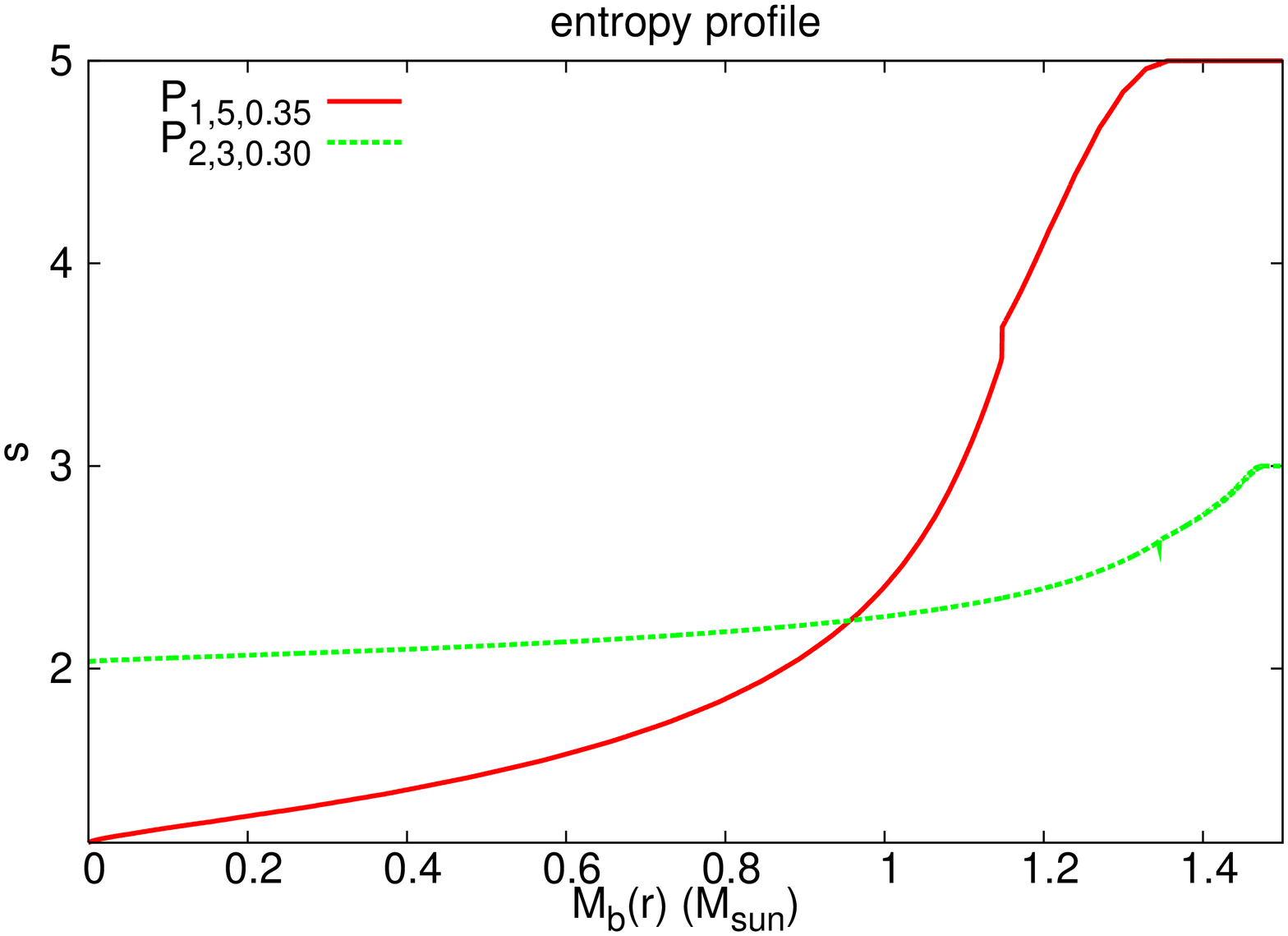,width=170pt,angle=270}
\epsfig{file=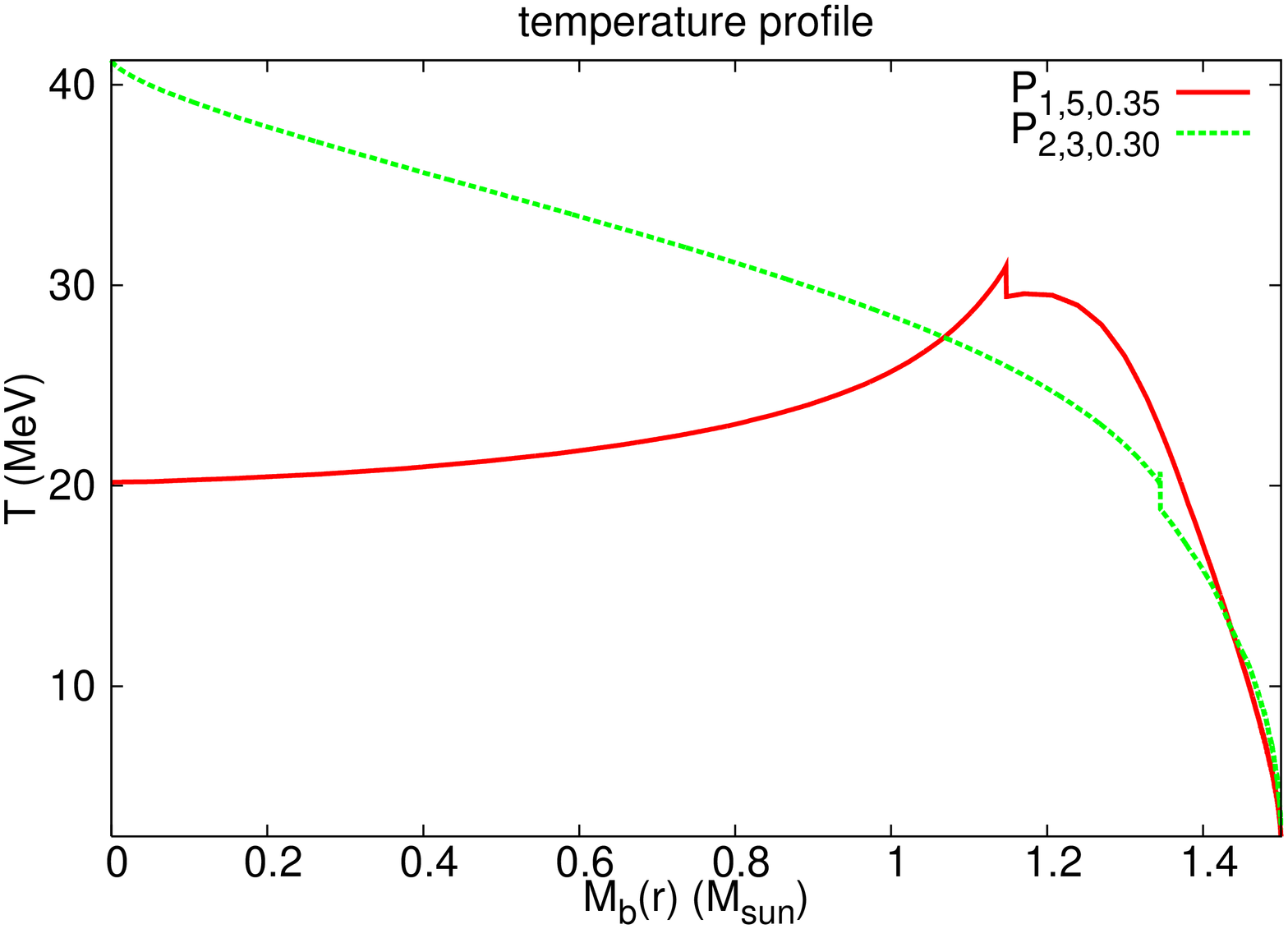,width=170pt,angle=270}
\caption{(Color online) 
The profiles of entropy (upper panel) and temperature (lower panel)
are plotted versus the enclosed baryonic mass for the models
$P_{1,5,0.35}$ ($R=30.3$ km) and $P_{2,3,0.30}$ ($R=16.5$ km).
}
\label{f:fig1}
\end{figure}

\begin{table}[t]
\caption{
Stellar models with fixed baryonic mass $M_B=1.5\,\msun$
corresponding to different entropy profiles and lepton fractions.
The gravitational mass $M$, radius $R$, central temperature $T^c$,
and central neutrino fraction $x_{\nu}^c$ are tabulated for each profile.
}
\vskip10pt
\begin{tabular}{|c|c|c||c|c|c|c|} 
\hline
$s_c$&$s_e$&$Y_e$&$M/\msun$&$R$ (km)&$T^c$ (MeV)&$x_\nu^c$\\
\hline
1.0&5.0&0.38&1.43&31.5&19.8&0.052\\
1.0&5.0&0.35&1.43&30.3&20.2&0.041\\ 
1.0&5.0&0.28&1.42&29.4&21.2&0.020\\  
1.5&4.5&0.33&1.43&24.6&30.3&0.035\\ 
1.5&4.5&0.32&1.42&24.4&30.5&0.031\\ 
2.0&4.0&0.32&1.43&21.5&40.5&0.033\\
2.0&4.0&0.30&1.43&21.3&41.0&0.027\\ 
2.0&3.0&0.30&1.42&16.5&41.2&0.026\\
2.0&3.0&0.28&1.41&16.4&41.8&0.021\\ 
2.0&2.0&0.28&1.41&14.5&41.6&0.020\\
2.0&2.0&0.23&1.40&14.1&42.9&0.010\\ 
1.0&1.0&0.23&1.37&12.5&20.2&0.007\\
\hline
1&1&$x_\nu=0$&1.36&12.2&20.9&0.000\\ 
\hline
\multicolumn{3}{|c||}{$T=0$}&1.35&11.9&0.00&0.000\\
\hline 
\end{tabular}
\label{t:profiles}
\end{table}

To describe the latest stages of the PNS
evolution we also consider two constant entropy profiles:
(i) one with $s_c=s_e=1$ and no neutrino
trapping, $P_{1,1,x_\nu=0}$,
with $Y_e$ varying from 0.10 at the center to 0.44 at the stellar
surface; (ii) a zero-temperature profile with no neutrino trapping,
$P_{T=0}$, which  describes a cold, old NS,  with  electron
fraction varying from 0.09 at the center to 0.44 at the surface.

We show in Fig.~\ref{f:fig1}  the profiles of the
entropy per baryon (upper panel) and of the temperature (lower panel)
as a function of the enclosed baryonic mass, for the models
$P_{1,5,0.35}$ (radius $R=30.3$ km) and $P_{2,3,0.30}$ (radius $R=16.5$
km).  In order to avoid sharp transitions from the core to the
envelope region, we adopt a cubic interpolation for the entropy
between the two regions.  Thus, the entropy is a continuous function of
the density. However, as a consequence of the Maxwell
construction used to join the Shen EOS (low-density region) to
the BHF EOS (high-density region), as discussed in the previous
section, there is  a weak discontinuity both in the
entropy  and in the temperature profile, when plotted as a function of
the enclosed mass as in Fig.~\ref{f:fig1}.
We have checked that the results presented in the next
section are not influenced by these discontinuities.

In Table \ref{t:profiles} we show the quantities which characterize
the stellar models associated to different profiles, 
namely  gravitational mass, stellar radius,
temperature and neutrino fraction at the center of the star.  The
dependence of the stellar parameters on the temperature and lepton
fraction profiles will be discussed in the next Section.

\section{Results}
\label{s:res}

In this section we discuss the behaviour of the stellar radius and of 
frequencies and damping
times of the QNMs computed for stellar models with different entropy
profiles and lepton/neutrino fraction content, in order to understand
how these quantities are affected by the PNS internal structure.

Let us consider the dependence on the entropy profile first, and
fix the value of the lepton fraction to
$Y_e=0.32$. We compute and compare  the mode frequencies and damping times of
the following stellar configurations: $P_{1,5,0.32}$, $P_{1,4,0.32}$,
$P_{1.5,4.5,0.32}$, $P_{2,4,0.32}$, and $P_{2,3,0.32}$.  The
core-envelope `entropy jumps' are $\Delta s=4,3,3,2,1$, respectively.
In Table \ref{t:tab2} we show for each profile the central
temperature, the radius, and the frequencies and damping times of the
QNMs $g_1,f,p_1$.  These data allow us to discuss how the different
quantities change with the entropy profile.

\begin{table}[t]
\caption{
Frequencies (in Hz) and damping times (in s) of the QNMs $g_1,f,p_1$ 
for different stellar models with baryonic mass $1.5\;\msun$,
lepton fraction $Y_e=0.32$, and different entropy profiles.
The central temperature $T^c$ (in MeV) and the stellar radius $R$ (in km)
are also shown.}
\medskip
\begin{tabular}{|c|c|c|c|c|c|c|c|c|c|c|c|}
\hline
$s_c$& $s_e$& $\Delta s$& $T^c$ & $R$ &$\nu_{g_1}$& $\tau_{g_1}$& 
$\nu_f$&$\tau_f$&$\nu_{p_1}$& $\tau_{p_ 1}$\\
\hline
1.0&5.0&4  &20.6&  29.6   & 906  & 6.27 & 1194 & 4.42 & 1528 & 0.75\\
1.5&4.5&3  &30.5&  24.3   & 910  & 42.9  & 1346 & 0.76 & 1845 & 0.55\\ 
1.0&4.0&3  &20.2&  18.4   & 870  & 793  & 1741 & 0.27 & 2574 & 0.99\\ 
2.0&4.0&2  &40.5&  21.5   & 669  & 2$\times10^3$ & 1449 & 0.45 & 2097 & 0.72\\
2.0&3.0&1  &40.7&  16.8   & 492  & 6$\times10^5$ & 1714 & 0.25 & 2977 & 1.64\\
\hline
\end{tabular}
\label{t:tab2}
\end{table}

As a general rule, the radius  is larger if the star is
hotter, or equivalently, if it has a larger entropy per baryon.  This
is indeed confirmed comparing for example the profiles $P_{1,5,0.32}$
and $P_{1,4,0.32}$.  The temperature (entropy) at the center is the
same, but the first model has larger entropy and temperature in the
envelope; its radius, $R=29.6$ km, is larger than that of the second
model, $R=18.4$ km.  This behaviour is confirmed by comparing
$P_{2,4,0.32}$ and $P_{2,3,0.32}$.  In a similar way, the radius
depends on the entropy in the core, although the dependence is weaker,
because the envelope has more freedom to expand than the core; for
instance, the configuration $P_{2,4,0.32}$ has a radius $R=21.5$ km,
larger than $R=18.4$ km of $P_{1,4,0.32}$.

As discussed in Section \ref{s:gmodes}, the frequency of the first \g
depends mainly on the core-envelope entropy jump: higher values of
$\Delta s=s_e-s_c$ correspond to larger \g frequencies.  Furthermore,
as argued in \cite{MVH88}, the \g frequency has also a (weaker)
dependence on the central temperature; indeed, the configurations
$P_{1.5,4.5,0.32}$ and $P_{1,4,0.32}$ have the same entropy jump, but
the former has a larger central temperature $T^c$ and larger \g
frequency.

\begin{figure}[t]
\epsfig{file=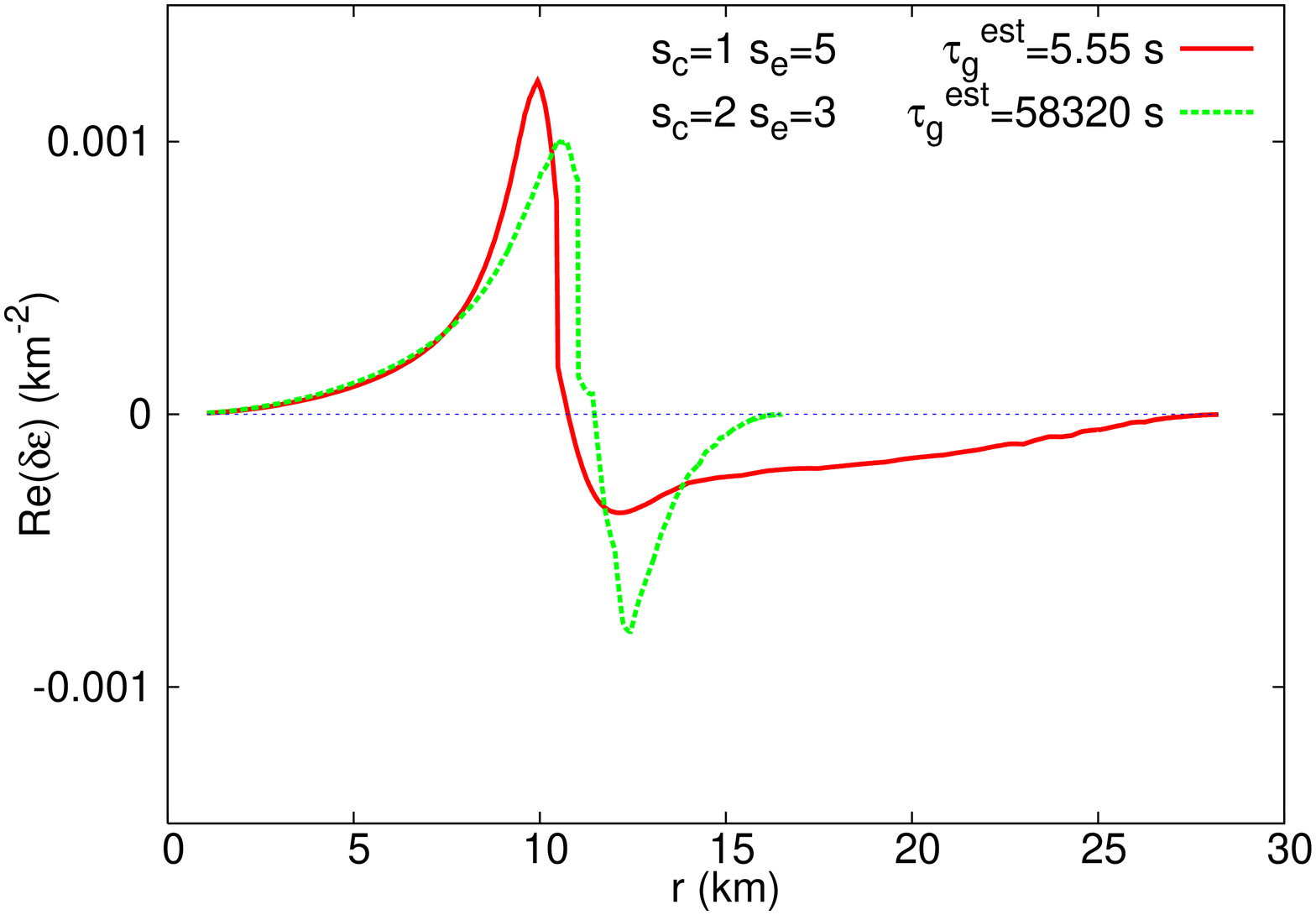,width=170pt,angle=270}
\epsfig{file=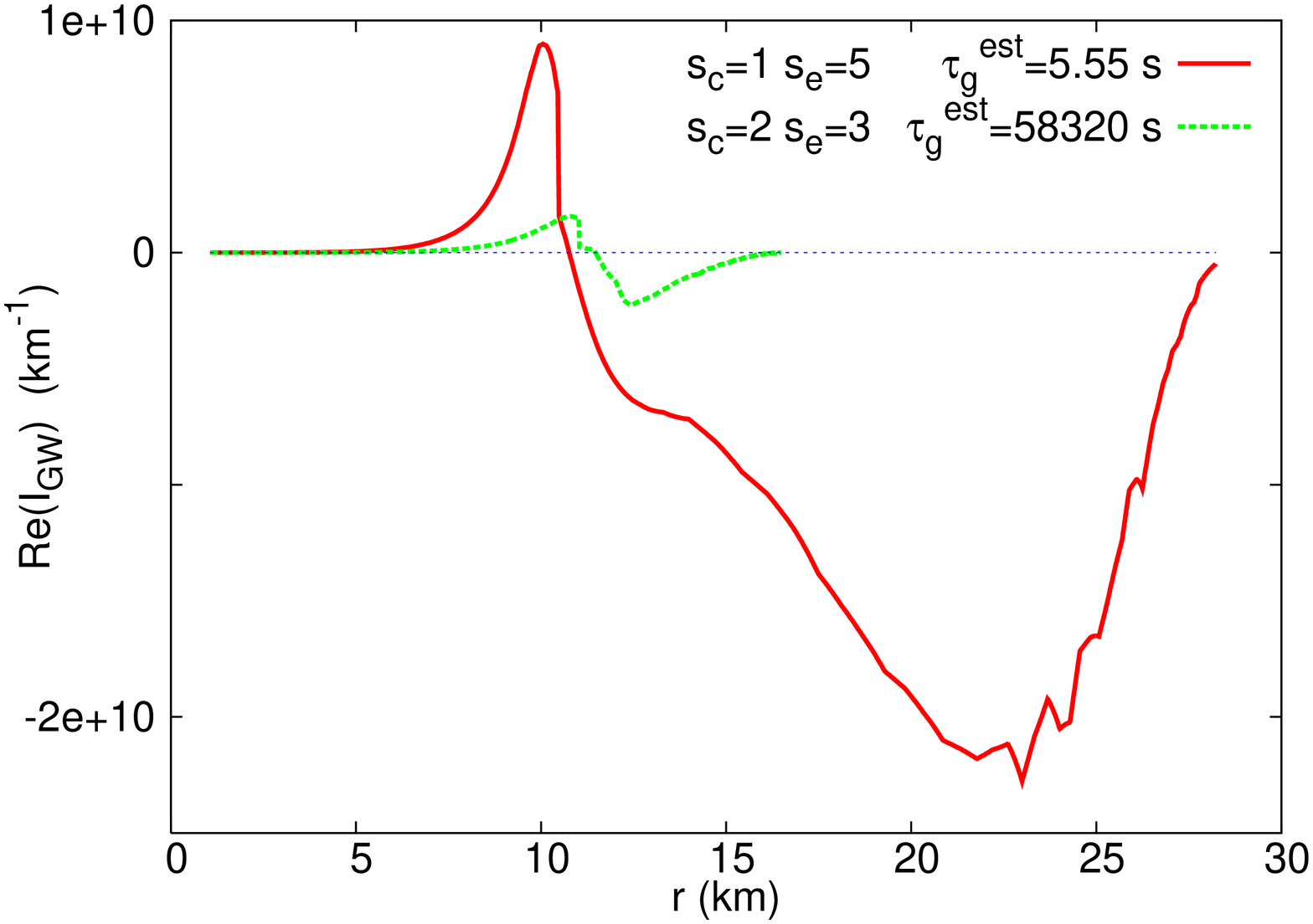,width=170pt,angle=270}
\caption{(Color online)  
Comparison of the perturbed energy density $\delta\eps(r)$, 
Eq.~(\ref{e:deltaeps}), (upper panel) 
and of the function $I_\text{GW}(r)$ given in Eq.~(\ref{IGW}) (lower panel) 
for the $g$-mode of the stellar configurations
$P_{1,5,0.32}$ and $P_{2,3,0.32}$.  
Both functions have been normalized in such a way that the mode pulsation 
energy is $E_\text{puls}=1$ km.}
\label{f:fig2}
\end{figure}

Table \ref{t:tab2} shows that, 
as the entropy jump decreases, the damping time of the first \g
increases dramatically: for $P_{1,5,0.32}$ it is $\tau_{g_1}=6.27$~s,
for $P_{2,3,0.32}$ it is $\tau_{g_1}=6\times 10^5$~s.  Since the
evaluation of the damping time is more sensitive to the numerical
procedure than that of the mode frequency, we have also computed 
this quantity
by using the approximate formula given in Eq.~(\ref{e:estimtau}), and indicate
its value as $\tau_g^\text{est}$;
we  find a reasonable agreement with the data
of Table~\ref{t:tab2}:
\begin{eqnarray*}
  \Delta s\!=\!4\!:~ &E_\text{puls}\!=\!1.047\, {\rm km},~ & 
  \lgw\!=\!0.377            ,~\tau_g^\text{est}\!=\!5.55\,{\rm s}
\\
  \Delta s\!=\!1\!:~ &E_\text{puls}\!=\!0.111\, {\rm km},~ & 
  \lgw\!=\!3.8\times 10^{-6},~\tau_g^\text{est}\!=\!6\!\times\!10^4\,{\rm s}\,.
\end{eqnarray*}
Thus, the sharp increase of $\tau_{g_1}$ as $\Delta s$ decreases is due to
the sharp decrease of the gravitational luminosity $\lgw$ [see
Eq.~(\ref{e:estimtau})]. As shown in Eq.~(\ref{LGW}), $\lgw$ is (modulo
a numerical factor) the squared integral of the function
\begin{equation}
  I_\text{GW} = \nu^3 r^4 \delta\eps^\lm \:,
\label{IGW}
\end{equation}
where $\nu$ is the mode frequency and $\delta\eps$ is the perturbed
energy density for the considered mode.

In order to understand why $\lgw$ decreases so much when the entropy
jump decreases, we plot in Fig.~\ref{f:fig2} $\delta\eps$ (upper
panel) and $I_\text{GW}$ (lower panel) as functions of $r$, for the
stars with profiles $P_{1,5,0.32}$ and $P_{2,3,0.32}$.  It is obvious
that due to the larger radius of the former configuration and the
presence of the factor $r^4$ in Eq.~(\ref{LGW}), the emitted power
$\lgw$ is much larger (indeed the main contribution comes from the
envelope), and the damping time is strongly reduced.  In addition, the \g
frequency is also larger in the former configuration, and this
contributes further to a larger gravitational wave emission, since
$I_\text{GW}\sim\nu^3$.

Let us now consider the $f$-modes.  For a cold neutron star, the \f
frequency scales as the average density of the star
and the damping time scales as $\tau_f\sim R^4/M^3$.
From the data of Table~\ref{t:tab2} we see that $\nu_f$ increases as the radius
decreases (the gravitational mass is nearly the same for all
configurations), while the damping time decreases; however, both $\nu_f$ 
and $\tau_f$ do not follow quantitatively  the cold star scaling laws. 
The first $p$-mode frequency has a behaviour similar to that of the
$f$-mode, whereas the damping time seems to be quite insensitive to 
changes of the entropy profile.


\begin{table}[t]
  \caption{
    Frequencies (in Hz) and damping times (in s) of the QNMs $g_1,f,p_1$ for stellar
    models with baryonic mass $1.5\;\msun$,
    entropy per baryon in the core $s_c=1$ and in the envelope $s_e=5$,
    and different values of the lepton fraction $Y_e$.
    The radius of the star (in km) and its gravitational mass (in
    solar masses $M_\odot$) are also shown.
  }
\medskip
\begin{tabular}{|c||c|c|c|c|c|c|c|c|}
\hline
$Y_e$ & $R$ & $M$ & $\nu_{g_1}$ & $\tau_{g_1}$ & 
$\nu_f$ & $\tau_f$ & $\nu_{p_1}$ & $\tau_{p_1}$\\
\hline
0.38 &  31.5 & 1.43  & 863  & 6.78 & 1116 & 9.75 &1415& 1.00\\ 
0.36 &  30.6 & 1.43  & 883  & 6.62 & 1147 & 6.83 & 1463 & 0.89\\
0.32 &  29.6 & 1.42  & 906  & 6.25 & 1194 & 4.44 & 1527 &  0.75\\
0.30 &  29.4 & 1.42  & 910  & 5.99 & 1209 & 4.01 & 1543 & 0.73\\
0.28 &  29.4 & 1.42  & 908  & 5.71 & 1216 & 3.96 & 1546 & 0.72\\
\hline
\end{tabular}
\label{table3}
\end{table}

Finally, we consider a sequence of stellar models with a fixed entropy
profile, i.e., $s_c=1$ in the core and $s_e=5$ in the envelope, and
lepton fraction varying in the range $Y_e = 0.38,\ldots,0.28$.  The
frequency and the damping times of the $g_1$, $f$, and $p_1$-modes are
shown in Table~\ref{table3}, together with the radius and the
gravitational mass of the star.  From these data we see that the star
radius is a slightly decreasing function of the lepton fraction, and
that the behaviour of the $f$- and $p_{1}$- frequency as a function of
the star radius is similar to that described above. Overall, the data
show that the dependence of the QNMs eigenfrequencies on the lepton
fraction is much weaker than that on the entropy profile.

\subsection{QNM eigenfrequencies  and  PNS quasi-stationary evolution}
As mentioned in the Introduction, numerical simulations show that in
the early phases of a PNS life the entropy profile has a
characteristic evolution which mainly depends on neutrino diffusion
processes, and which can be divided in three essential steps:
\begin{enumerate}
\item the entropy per baryon is initially (a few tenths of seconds after
bounce) larger in the envelope and lower in the core;
\item the entropy increases in the core and decreases in the envelope,
reaching a roughly uniform profile;
\item  the entropy decreases throughout the star,
which eventually becomes a cold neutron star.
\end{enumerate}

The entire process takes about a minute, but we cannot assign precise
temporal labels to each step, because they depend on the details of
the initial conditions after the bounce and on the dynamical modeling
of the evolution, which is beyond the scope of our work (an example of
this evolution is shown in Fig.~9 of Ref.~\cite{pons1}).

In this section we construct a sequence of stellar configurations, listed in
Table~\ref{t:tab4}, which captures the main qualitative features of a
PNS evolution described by steps 1 to 3.  Each profile is labeled by a
number $i$, which gives the ordering in time of the simulated
evolution.  Configurations from $i=1$ to $i=4$ (envelope entropy
larger than core entropy) refer to the transition from step 1 to step
2, which ends with configuration 5, for which the entropy distribution becomes
uniform, but the star is still hot.  Then it cools down
(configurations 5 to 6) and ends as a zero-temperature NS reaching
configuration 7 (step 3).  During this ``evolution'' the lepton number
decreases.  We have also considered a different sequence, in which the
lepton fraction decreases ``more rapidly,'' but the results are very
similar to those obtained with the sequence shown in
Table~\ref{t:tab4}.

\begin{table}[t]
\caption{
Frequencies (in Hz) and damping times (in s) of the QNMs $g_1,f,p_1$
for a sequence of stellar models which mimic the
quasi-stationary evolution
of a PNS  with constant baryonic mass $1.5\;\msun$.
The star radius (in km) is shown in column 5.
}
\medskip
\begin{tabular}{|c|c|c|c||c|c|c|c|c|c|c|}
\hline
$i$&$s_c$&$s_e$&$Y_e$&$R$ & $\nu_{g_1}$& $\tau_{g_1}$& $\nu_f$&$\tau_f$&
$\nu_{p_1}$& $\tau_{p_1}
$\\
\hline
1&1.0&5.0&0.35&30.3&890&6.54&1162&5.89&1484&0.84\\
2&1.5&4.5&0.32&24.4&910&42.9&1346&0.76&1845&0.55\\
3&2.0&4.0&0.30&21.3&667&2.3$\times 10^3$&1452&0.44&2125&0.73\\
4&2.0&3.0&0.28&16.4&485&7.6$\times 10^4$&1717&0.25&3133&1.80\\
5&2.0&2.0&0.23&14.1&0&-&1790&0.23&4134&2.59\\
\hline
6&1.0&1.0&$x_\nu=0$&12.2&0&-&1896&0.21&5879&2.98\\
\hline
7&\multicolumn{3}{|c||}{$T=0$}&11.9&0&-&1898&0.21&6006&3.52\\
\hline
\end{tabular}
\label{t:tab4}
\end{table}

\begin{figure}[t]
\epsfig{file=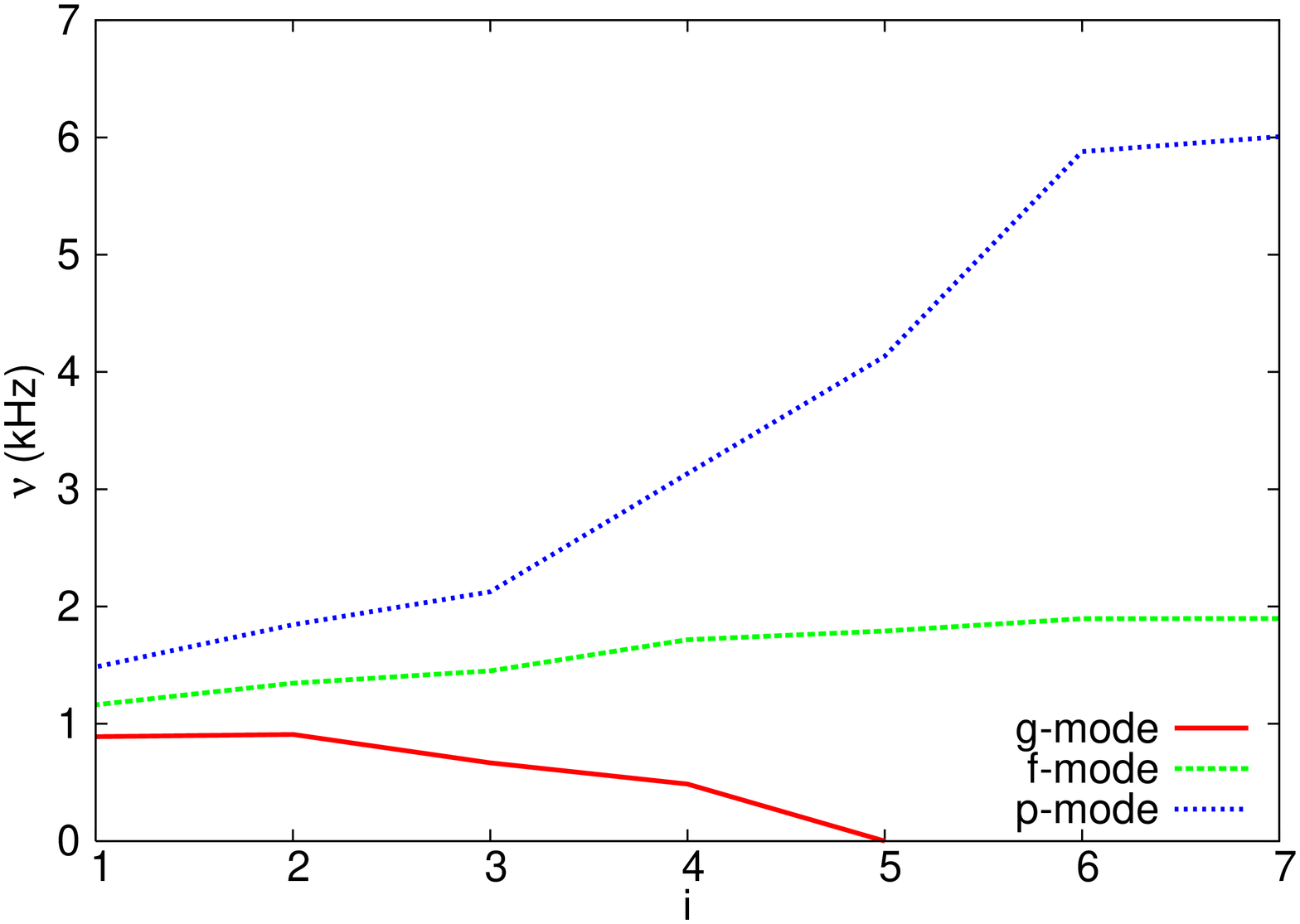,width=170pt,angle=270}
\epsfig{file=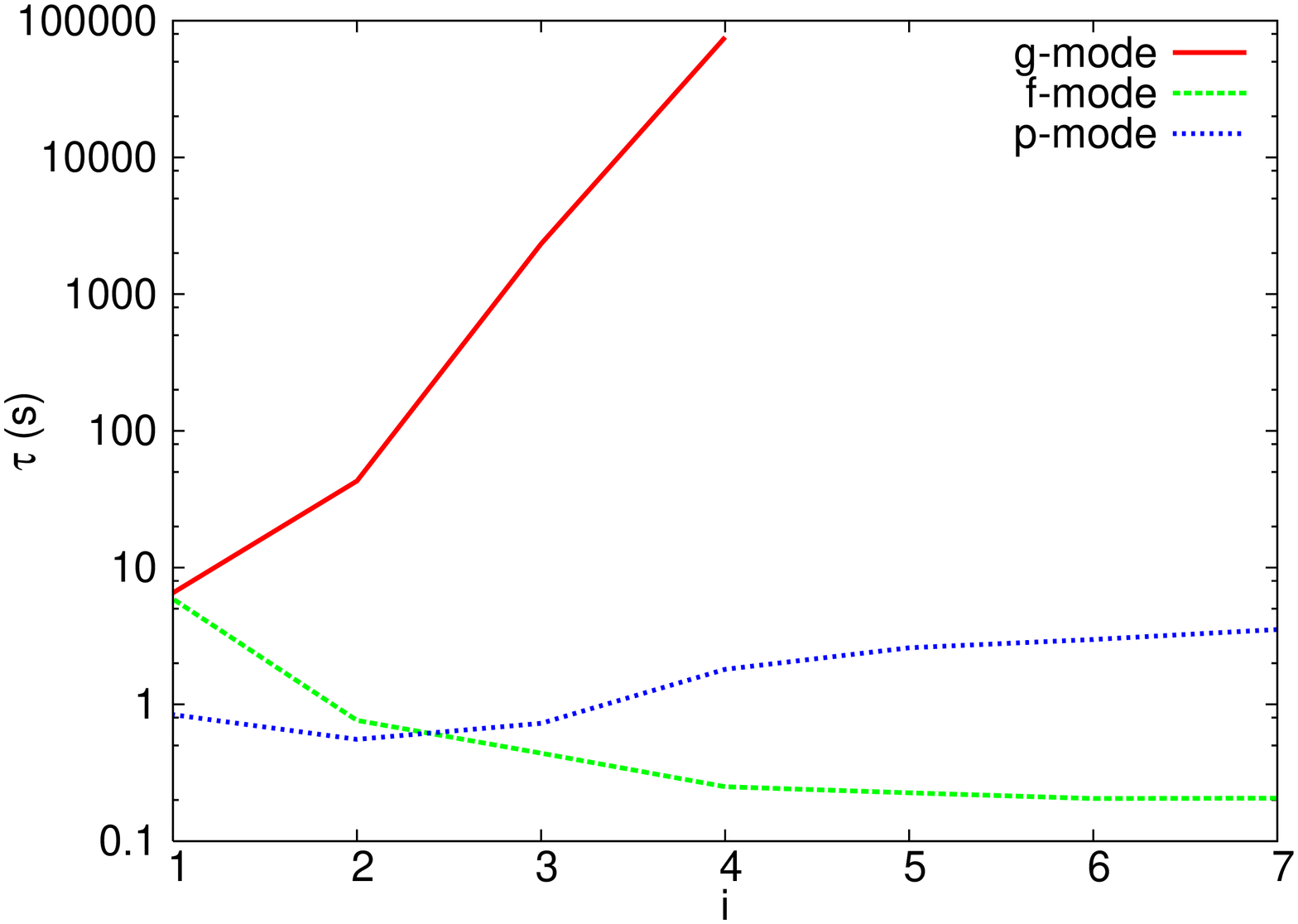,width=170pt,angle=270}
\caption{(Color online) 
Frequencies (upper panel) and damping times (lower panel)
of the QNMs $g_1,f,p_1$ for stellar models corresponding
to a possible evolutive sequence of stationary configurations.}
\label{f:fig3}
\end{figure}

For each configuration we compute  the
frequencies and damping times of the QNMs $g_1,f,p_1$.  Their values
are given in Table~\ref{t:tab4}, and are plotted in Fig.~\ref{f:fig3}
versus the number $i$ which identifies the configuration as explained
above. We remark that, as shown in Fig.~\ref{f:fig3}, the
gravitational damping time of the $g_1$-mode sharply increases for
$i\gtrsim 2$, while the mode frequency sharply decreases.
However, as discussed in Sec.~\ref{s:osc}, as soon as $\tau_g$ becomes
comparable to $\tau_\text{diss}\sim10-20$ s, the mode becomes
ineffective with respect to gravitational wave emission, since
the stellar oscillations are damped by non-gravitational dissipative processes.

The most interesting result which emerges from Table~\ref{t:tab4} and
Fig.~\ref{f:fig3} is that at earlier times, i.e., for $i \leq 2$, the
frequencies of the $g_1,f,p_1$ modes cluster in a small region around
1~kHz, and then tend quite rapidly to the values appropriate for a
cold NS (remember that our entire sequence should cover approximately
a minute of the PNS evolution).  This behaviour is similar to that
found in \cite{ferrari_miniutti_pons_2003}, where the quasi-stationary
evolution sequence was obtained using a finite-temperature EOS derived
within the mean field approach, treating neutrino transport using the
diffusion approximation.  The fact that in the very early stages the
mode frequency is of the order of 1 kHz  (or lower)
is important for gravitational wave detection,
because the sensitivity of ground-based
interferometers LIGO/VIRGO decreases quite significantly at larger
frequencies.

Another interesting point to note is that during the early evolution
the damping time of all modes is smaller than 10~s.  This means that
gravitational wave emission is effectively competing in removing energy from the star
with dissipation processes related to neutrino
viscosity, diffusivity, and thermal conductivity, since typical
neutrino timescales are of the order of 10--20~s (see also section 2.1 of
\cite{ferrari_miniutti_pons_2003} for a detailed discussion of this
point).

Furthermore, for $i=1$ the damping times of the \f and \g are nearly
coincident, showing that in the early stages the \g is as effective as the \f as a source
of gravitational waves.  Indeed, the function $I_\text{GW}$ given by
Eq.~(\ref{IGW}), whose square integral over the star is the
gravitational wave luminosity, is similar for the two modes due to a
similar profile of the energy density perturbation.  At later
``times'' $\tau_g$ becomes much larger than $\tau_f$ and,
as the PNS tends to the NS final configuration,
the \g frequency tends to zero.

\section{Conclusions}
\label{s:end}

In this paper we have investigated how the frequencies and damping
times of the quasi-normal modes of a proto-neutron star depend on the
physical quantities which characterize the stellar configurations
during the quasi-stationary evolution.  The most important is the
entropy profile inside the star, whereas the dependence on the lepton
composition is weaker.

The most interesting result is that if the entropy gradient between core
and envelope is large, the frequencies of the first $g$-mode, of the
fundamental mode, and of the first $p$-mode tend to cluster in a small
region near $1$ kHz, whereas the damping time of the first $g$-mode
and of the $f$-mode become comparable.  This means that during the
initial phases of the quasi-stationary evolution, when the core
entropy is low and the envelope entropy is large, these two modes are
competitive as far as gravitational wave emission is concerned.

The damping times are of the order of a few seconds, smaller than
dissipative timescales associated to neutrino processes, which are of
the order of 10--20~s.  Thus, if the star has some mechanical energy
to dissipate, it is likely that it will do it 
through  the $g_1$ and $f$ modes.

\section*{Acknowledgements}
We thank J.A. Pons, K. Glampedakis and D.I. Jones for useful
discussions.  This work was partially supported by CompStar, a
Research Networking Programme of the European Science Foundation, and
by the MIUR-PRIN Project 2008KRBZTR. L.G. has been partially supported
by the grant PTDC/FIS/098025/2008.

\appendix
\begin{widetext}
\section{The Lindblom-Detweiler equations}\label{LD}

The system of the LD equations \cite{LD83,DL85} consists of four first-order
differential equations in the quantities
$H_1^\lm(r)$, $K^\lm(r)$, $W^\lm(r)$, $X^\lm(r)$:
\begin{eqnarray}
 H_1^{lm\,\prime} &=&
 -\frac{1}{r} \biggl[ \ell+1+\frac{2Me^\lambda}{r}+4\pi
  r^2e^\lambda(p-\epsilon) \biggr]
 + \frac{e^\lambda}{r}
 \left[ H_0^\lm + K^\lm - 16\pi(\epsilon+p)V^\lm \right] \:,
\nonumber\\
 K^{\lm\,\prime} &=&
 \frac{1}{r} H_0^\lm + \frac{\ell(\ell+1)}{2r} H_1^\lm
 - \left[ \frac{\ell+1}{r}+\frac{\psi'}{2} \right] K^\lm 
 - 8\pi(\epsilon+p)\frac{e^{\lambda/2}}{r} W^\lm \:,
\nonumber\\
 W^{\lm\,\prime} &=&
 - \frac{\ell+1}{r} W^\lm 
 + re^{\lambda/2} \left[ \frac{e^{-\psi/2}}{(\epsilon+p)c_s^2} X^\lm 
 - \frac{\ell(\ell+1)}{r^2} V^\lm + \frac{1}{2}H_0^\lm + K^\lm \right] \:,
\nonumber\\
 X^{\lm\,\prime} &=&
 -\frac{\ell}{r} X^\lm + \frac{(\epsilon+p)e^{\psi/2}}{2}
 \Biggl[ \left( \frac{1}{r}+\frac{\psi'}{2} \right)
 + \left(r\omega^2e^{-\psi} + \frac{\ell(\ell+1)}{2r}\right) H_1^\lm
 + \left(\frac{3}{2}\psi' - \frac{1}{r}\right) K^\lm
\nonumber\\&&
 - \frac{\ell(\ell+1)}{r^2}\psi' V^\lm 
 - \frac{2}{r} 
 \Biggl( 4\pi(\epsilon+p)e^{\lambda/2} + \omega^2e^{\lambda/2-\psi}
 - \frac{r^2}{2}
 \biggl(\frac{e^{-\lambda/2}}{r^2}\psi'\biggr)' \Biggr) W^\lm \Biggr] \:.
\label{LLDD}
\end{eqnarray}
The remaining perturbation functions,
$H_0^\lm(r)$, $V^\lm(r)$, $H_2^\lm(r)$,
are given by the algebraic relations
\begin{eqnarray}
 0 &=&
 \left[ 3M + \frac{(\ell-1)(\ell+2)}{2}r + 4\pi r^3p \right] H_0^\lm
 - 8\pi r^3e^{-\psi/2} X^\lm
 + \left[ \frac{\ell(\ell+1)}{2}(M+4\pi r^3p) 
   - \omega^2r^3e^{-(\lambda+\psi)} \right] H_1^\lm
\nonumber\\&&
 - \left[ \frac{(\ell-1)(\ell+2)}{2}r - \omega^2r^3e^{-\psi}
 + \frac{e^\lambda}{r}(M+4\pi r^3p)(3M-r+4\pi r^3p) \right] K^\lm \:,
\nonumber\\
 X^\lm &=&  
 \omega^2(\epsilon+p)e^{-\psi/2} V^\lm
 - \frac{p'}{r}r^{(\psi-\lambda)/2} W^\lm
 + \frac{e^{\psi/2}}{2}(\epsilon+p) H_0^\lm \:,
\nonumber\\
 H_0^\lm &=& H_2^\lm \:.
\label{LDf}
\end{eqnarray}

Equations (\ref{LLDD}) and (\ref{LDf}) are solved numerically inside the star,
assuming that the perturbation functions are non-singular near the center.
An asymptotic expansion of the equations 
near $r=0$ shows that this requirement implies
\begin{eqnarray}
 X^\lm(0) &=& 
 \bigl[ \epsilon(0)+p(0)\bigr] e^{\psi(0)/2} 
 \left[ \biggl( \frac{4\pi}{3} \bigl[\epsilon(0)+3p(0)\bigr] 
 - \omega^2\frac{e^{-\psi(0)}}{\ell} \biggr) W^\lm(0)
 + \frac{1}{2}K^\lm(0) \right] \:,
\nonumber\\
 H_1^\lm(0) &=&
 \frac{1}{\ell(\ell+1)} \left[ 2\ell K^\lm(0)
 + 16\pi \bigl[\epsilon(0)+p(0)\bigr] W^\lm(0) \right] \:.
\label{bc1}
\end{eqnarray}
\end{widetext}

On the stellar surface, $r=R$, one assumes continuity of the perturbation
functions and the vanishing of the Lagrangian pressure perturbation, i.e.,
\begin{equation}
 X^\lm(R) = 0 \:.
\label{bc2}
\end{equation}
In the exterior, the metric perturbations are described by the Zerilli function 
\begin{equation}
 Z^\lm = \frac{r^{\ell+2}}{nr+3M} \left( K^\lm - e^\psi H_1^\lm \right) \:,
\end{equation}
[where $n=(\ell-1)(\ell+2)/2$], which is solution of the Zerilli equation
\begin{equation}
 \frac{d^2Z^\lm}{dr_*^2} + \left[ \omega^2 - V_Z(r) \right] Z^\lm
 = 0
\label{zereq}
\end{equation}
with $r_* \equiv r + 2M\ln(r/2M-1)$ and
\begin{equation}
 V_Z \equiv e^{-\lambda} 
 \frac{2n^2(n+1)r^3+6n^2Mr^2+18nM^2r+18M^3}{r^3(nr+3M)^2} \:.
\end{equation}

Finally, to describe free oscillations of the star we must impose the 
outgoing wave boundary condition
\begin{equation}
 Z^\lm(r) \rightarrow e^{-\ii\omega r_*}
 ~~~(r\rightarrow\infty) \:.
\label{bc3}
\end{equation}
A solution of Eqs.~(\ref{LLDD}), (\ref{zereq})
satisfying the boundary conditions (\ref{bc1}), (\ref{bc2}), (\ref{bc3})
only exists for a discrete set of (complex) values of the frequency
$\omega = 2\pi\nu + \ii/\tau$:
the quasi-normal modes of the star.


\end{document}